\documentstyle{l-aa}
\input{epsf}

\def\la{\;
\raise0.3ex\hbox{$<$\kern-0.75em\raise-1.1ex\hbox{$\sim$}}\; }
\def\ga{\;
\raise0.3ex\hbox{$>$\kern-0.75em\raise-1.1ex\hbox{$\sim$}}\; }

\begin{document}

\title{ Neutrino synchrotron emission from dense
        magnetized electron gas of neutron stars}
\author{V.G.~Bezchastnov$^1$, P.~Haensel$^2$, A.D.~Kaminker$^1$,
        D.G.~Yakovlev$^1$}
\offprints{P.~Haensel}
\institute{A.F.~Ioffe Physical Technical Institute,
              194021 St.Petersburg, Russia
              \and
              N.~Copernicus Astronomical Center,
              Polish Academy of Sciences,
              Bartycka 18, 00-716 Warszawa, Poland}
\thesaurus{02.04.1 - 08.09.3 - 08.14.1}
\date{}
\maketitle
\markboth{V.G.\ Bezchastnov et al.: Neutrino synchrotron emission \ldots}{}
\label{sampout}

\begin{abstract}
We study the synchrotron emission of neutrino pairs by
relativistic, degenerate electrons in strong
magnetic fields. Particular attention is paid to
the case in which the dominant contribution comes
from one or several lowest cyclotron
harmonics. Calculations are performed using
the exact quantum formalism and the quasiclassical
approach. Simple analytic fits to the neutrino synchrotron
emissivity are obtained in the domain
of magnetized, degenerate and relativistic electron gas
provided the electrons populate either many Landau levels
or the ground level alone.
The significance of the neutrino synchrotron energy
losses in the interiors of cooling neutron stars is discussed.
\end{abstract}

\section{Introduction}
 In this article, we study the synchrotron emission of neutrino
pairs by relativistic, degenerate electrons in strong
magnetic fields,
\begin{equation}
     e \rightarrow e + \nu + \bar{\nu}.
\label{synchro}
\end{equation}
This emission can be an important source of neutrino
energy losses from the crusts  of magnetized cooling
neutron stars (NSs) as well as (in a modified form)
from the superfluid  NS cores.

The process has been considered by several authors
starting from the pioneering article by Landstreet (1967).
The adequate quasiclassical formalism was developed
by Kaminker, Levenfish \& Yakovlev (1991, hereafter KLY)
who presented also critical analysis of preceding work.
KLY carried out detailed analytic and numerical analysis of the
case in which the electrons populated many Landau levels
and the main contribution into the synchrotron emission came from
high cyclotron harmonics. They also considered briefly
the opposite case of the superstrong magnetic field
which suppresses greatly the contribution from all harmonics.
In that case, only the first harmonics actually survives
(although exponentially damped).

Exact quantum formalism of the neutrino
synchrotron emission was developed by Kaminker et al.\ (1992a).
Kaminker \& Yakovlev (1993) considered the synchrotron
process in the nondegenerate electron gas. Recently
Kaminker et al.\ (1997) have analyzed how the synchrotron
emission is modified in the superconducting NS cores
where the initially uniform magnetic field splits into fluxoids.
In a recent article, Vidaurre et al.\ (1995) have reconsidered the results
of KLY claiming them to be inaccurate.

In the present article,
we continue to study the neutrino synchrotron radiation
from the degenerate magnetized electron gas. The emphasis is
made upon the case in which
the electron transitions associated with
one or several lowest cyclotron
harmonics are most important. This case has not been studied attentively
earlier. We calculate the neutrino emissivity numerically
using the exact quantum 
formalism (Sect.\ 2, Appendix A). We
also use the quasiclassical approach (Sect.\ 3, Appendix B),
combine our new results with those by KLY
and propose an analytic fit that describes accurately all
the neutrino synchrotron radiation regimes in a strongly degenerate,
relativistic electron gas
in which the electrons occupy many Landau levels.
Another limiting case in which the electrons
populate the ground level alone is considered in Appendix C.
In Sect.\ 3 we present also critical discussion of the
results by Vidaurre et al.\ (1995). In Sect.\ 4 we
show the importance of the neutrino synchrotron
emission in the NS interiors.

\section{Quantum formalism}
We will mainly use the units in which $m_e=c=\hbar=k_{\rm B}=1$,
where $k_{\rm B}$ is the Boltzmann constant. We will return
to ordinary physical units whenever necessary.
The general expression for the neutrino synchrotron
energy loss rate (emissivity, ergs s$^{-1}$ cm$^{-3}$)
from an electron gas of any degeneracy and relativity,  
immersed in a quantizing magnetic field, 
was obtained by Kaminker et al.\ (1992a):
\newpage
\begin{eqnarray}
     Q_{\rm syn} &  = &  \frac{G_{\rm F}^2 b}{3 (2\pi)^5}
     \sum_{n=1,n'=0}^\infty
     \int_{-\infty}^\infty {\rm d} p_z
     \int {\rm d} q_z \int q_\perp {\rm d} q_\perp
\nonumber   \\
     &  \times &
     A \omega f(1-f').
\label{Quantum}
\end{eqnarray}
Here, $G_{\rm F}= 1.436 \times 10^{-49}$~ergs~cm$^{3}$ is the Fermi
weak-coupling constant and $b=B/B_c$ is
the dimensionless magnetic field ($B_c =
m_e^2c^3/(\hbar e) \approx 4.414 \times 10^{13}$~G).
Furthermore, $n$ and $p_z$ are, respectively, the
Landau level number and the momentum along the magnetic field
for an electron before a neutrino-pair emission;
the energy of this electron is $\varepsilon=\sqrt{1+2nb+p_z^2}$.
The primed quantities $n'$ and $p'_z$ refer to an electron
after the emission; its energy is
$\varepsilon'=\sqrt{1+2n'b+p_z^{\prime 2}}$;
$f= f(\varepsilon) = \{ \exp [(\varepsilon - \mu)/T]+1 \}^{-1}$
is Fermi-Dirac distribution of the initial-state electrons,
$f'=f(\varepsilon')$ is the same for the final-state electrons;
$\mu$ is the electron chemical potential and $T$ is the temperature.
The energy and momentum
carried away by a neutrino-pair are denoted as
$\omega = \varepsilon - \varepsilon'$ and {\bf q}, respectively.
The $z$ component of {\bf q} is $q_z=p_z -p'_z$,
while the component of {\bf q} across the magnetic field
is denoted by $q_\perp$.
The summation and integration in (\ref{Quantum})
is over all allowed electron transitions.
The integration has to be done over the
kinematically allowed domain
$q_z^2 + q_\perp^2 \leq \omega^2$.
The differential transition rate $A$
(summed over initial and final electron spin states)
is given by Eq.\ (17) of Kaminker et al.\ (1992a).
Taking into account that some terms
are odd functions of $p_z$ and vanish after the integration,
we can rewrite $A$ in a simple form,
\begin{eqnarray}
       A & = & \frac{C_{+}^2}{2 \varepsilon \varepsilon'}
              \left\{
              \left[
              \left( \omega^2 - q_z^2 - q_{\perp}^2   \right)^{}
              \left( p_{\perp}^2 + p_{\perp}^{\prime 2} + 2  \right)
              + q_{\perp}^2   \right]
              (\Psi - \Phi)  \right.
\nonumber    \\
             & - &
              \left.
             \left( \omega^2 - q_z^2 - q_{\perp}^2  \right)^2
                \Psi +   \left( \omega^2 - q_z^2 - q_{\perp}^2
             \right) \Phi   \right\}
\nonumber   \\
             & - &  \frac{C_{-}^2}{2 \varepsilon \varepsilon'}
             \left[
             \left( 2 \omega^2 - 2 q_z^2 - q_{\perp}^2  \right)
             (\Psi - \Phi)   \right.
\nonumber   \\
         &  + &  \left.
            3 \left( \omega^2 - q_z^2 -q_{\perp}^2  \right) \Phi
         \right].
\label{A_general}
\end{eqnarray}
Here, $p_\perp = \sqrt{2nb} \,$ and $p'_\perp = \sqrt{2n'b} \,$ are
the transverse momenta of the initial-state and final-state electrons,
respectively;
\begin{eqnarray}
    \Psi & = & F_{n'-1,n}^2(u) + F_{n',n-1}^2(u),
\nonumber  \\
    \Phi & = & F_{n'-1,n-1}^2(u) + F_{n',n}^2(u),
\label{PsiPhi}
\end{eqnarray}
$u = q_\perp^2 /(2b)$,
$F_{n'n}(u)=(-1)^{n'-n}F_{nn'}(u)=
u^{(n-n')/2} \, {\rm e}^{-u/2} \,( n'!/n! )^{1/2} \, L_{n'}^{n-n'}(u)$,
and $L_n^s(u)$ is an associated Laguerre polynomial
(see, e.g., Kaminker \& Yakovlev 1981).
Furthermore, in Eq.\ (\ref{A_general})
we introduce $C_+^2=\sum_\nu (C_V^2+C_A^2)$ and
$C_-^2=\sum_\nu (C_V^2-C_A^2)$, where
$C_V$ and $C_A$ are
the vector and axial vector
weak interaction constants, respectively,
and summation is over all neutrino flavors. For the
emission of the electron neutrinos
(charged + neutral currents), one has
$C_V = 2 \sin^2 \theta_{\rm W} +0.5$ and $C_A= 0.5$,
while for the emission of the muonic or tauonic
neutrinos (neutral currents only),
$C'_V = 2 \sin^2 \theta_{\rm W} - 0.5$ and $C'_A = -0.5$;
$\theta_{\rm W}$ is the Weinberg angle. Adopting
$\sin^2 \theta_{\rm W} \simeq 0.23$ we obtain
$C_+^2 \approx 1.675$ and $C_-^2 \approx 0.175$.
A comparison of the neutrino synchrotron emission by electrons
with the familiar electromagnetic synchrotron emission
has been done by KLY. Although electromagnetic radiation
is much more intense it does not emerge from deep neutron
star layers 
 (neutron star interior is opaque to photons) 
  while neutrino emission escapes freely from stellar
interior, producing an efficient internal cooling.

We have composed a computer code which calculates
$Q_{\rm syn}$ from Eqs.\ (\ref{Quantum}) and
(\ref{A_general}) for arbitrary plasma parameters
$\rho Y_e$,  $T$ and $B$,
where $\rho$ is the mass density and $Y_e$
is the number of electrons per baryon.
Technical details are presented
in Appendix~A, the results are illustrated in
Sect.\ 4, and the case of very strong magnetic field,
in which 
the bulk of electrons 
occupy the ground Landau level,
is considered in Appendix~C.

\section{Quasiclassical treatment}
In this section, we 
develop quasiclassical
description of the neutrino synchrotron 
emission
from a degenerate, ultrarelativistic electron gas
in a strong magnetic field $B= 10^{11}$--10$^{14}$~G.
Therefore we assume that $\mu \gg m_ec^2$ and $T \ll T_{\rm F}$,
where $T_{\rm F}=(\mu-m_e c^2)/k_{\rm B}
\approx \mu / k_{\rm B}$ is the degeneracy temperature.
We mainly analyze the case in which the electrons populate
many Landau levels. This is so if the magnetic field
is nonquantizing or weakly quantizing
(e.g., Kaminker \& Yakovlev, 1994), i.e., if $\mu \gg \sqrt{1 + 2b}$.
The latter condition is realized at sufficiently high densities
$\rho \gg \rho_B$, where
$\rho_B \approx 2.07 \times 10^6 b^{3/2}/Y_e$~g~cm$^{-3}
= 2.23 \times 10^5 B_{13}^{3/2}/Y_e$~g~cm$^{-3}$,
and $B_{13}=B/(10^{13}~{\rm G})$.
At these densities, $\mu$ is nearly the same as without magnetic field,
$\mu \approx c p_{\rm F} \approx c \hbar (3 \pi^2 n_e)^{1/3}$,
where $p_{\rm F}$ is the field--free Fermi momentum.
The formulated conditions are typical for the neutron star crusts.

In the case of many populated Landau levels, we can
replace the summation over $n$ by the integration over $p_\perp$
in Eq.\ (\ref{Quantum}). The remaining summation over $n'$ can be
conveniently replaced by the summation over
discrete cyclotron harmonics
$s=n-n'$=1, 2, 3, \ldots Furthermore,
an initial-state electron
can be described by its quasiclassical momentum
$p$ and pitch-angle $\theta$ ($p_z = p \cos \theta$,
$p_\perp = p \sin \theta$).
Since we consider strongly degenerate
electrons we can set $\varepsilon = \mu$ and
$p=p_{\rm F}$ in all smooth functions under the integrals.
Then the integration over $p$ is done analytically.
In this way we transform the rigorous quantum formalism of
Sect.\ 2 to
the quasiclassical approximation used by KLY (see their Eqs.\ (3),
(4) and (8)).

According to KLY
the quasiclassical neutrino
synchrotron emission is different in three temperature
domains ${\cal A}$, ${\cal B}$, and ${\cal C}$
separated by two typical temperatures $T_P$
and $T_B$:
\begin{eqnarray}
     T_P & = & {3 \hbar \omega_B^\ast x^3 \over 2 k_{\rm B}}
               ={3 \over 2} T_B x^3
               \approx 2.02 \times 10^9 B_{13} x^2~~{\rm K},
\nonumber \\
     T_B & = & {\hbar \omega_B^\ast \over k_{\rm B}}
         \approx 1.34 \times 10^9 B_{13} (1+x^2)^{-1/2}~~{\rm K}.
\label{T}
\end{eqnarray}
Here, $x=p_{\rm F}/(m_e c)$ is the relativistic parameter
($x \gg 1$, in our case),
and $\omega_B^\ast = eBc/\mu$ is
the electron gyrofrequency at the Fermi surface.
Notice, that while applying our results to the
neutron star crusts, one can use a simplified expression
$x \approx 1.009 (\rho_6\, Y_e)^{1/3}$, where
$\rho_6 = \rho/ (10^6~{\rm g~cm}^{-3})$.

\begin{figure}
 \hskip  1cm
 \epsfxsize=8.0cm
 \epsfbox{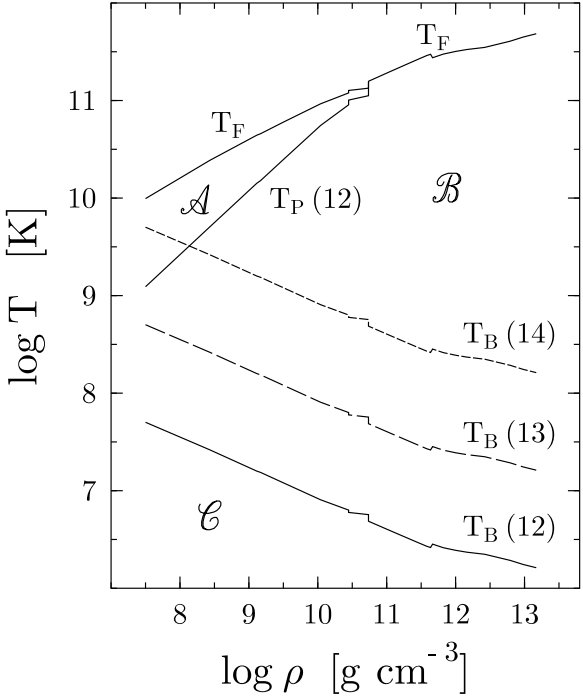}
\caption[ ]{
Density--temperature domains of different
neutrino synchrotron regimes.
$T_{\rm B}$ is temperature  \protect(\ref{T})
below which (domain ${\cal C}$)
the emission goes via the ground cyclotron
harmonics $s=1$ (for $B=10^{12}$, 10$^{13}$ and 10$^{14}$~G,
$\log B$ shown in parenthesis). $T_{\rm P}$ is temperature
\protect(\ref{T})
below which (in domain ${\cal B}$)
the Pauli principle 
restricts 
the number of cyclotron harmonics (see text);
it is shown only for $B=10^{12}$~G because
$T_{\rm P}$ becomes higher than
the electron degeneracy temperature
$T_{\rm F}$  for  $B=10^{13}$ and $10^{14}$~G.
Domain ${\cal A}$ ($T_{\rm P} \ll T \ll T_{\rm F}$)
is realized only for $B=10^{12}$~G;
$T_{\rm F}$ is independent of $B$, for displayed parameters.
}
\label{fig1}
\end{figure}

Figure 1 demonstrates the main parameter domains
${\cal A}$, ${\cal B}$ and ${\cal C}$ for the ground--state
(cold catalyzed) matter of NS crusts.
Thermal effects on the
nuclear composition are neglected which is justified as long as
$T \la 5 \times 10^9$ K (e.g., Haensel et al.\ 1996).
It is assumed
that nuclei of one species are available at any fixed density
(pressure). Then the increase
of density (pressure) is accompanied by
jumps of nuclear composition (e.g., Haensel et al.\ 1996).
The ground--state matter in the outer NS crust, at densities below
the neutron drip density ($4 \times 10^{11}$ g cm$^{-3}$), is described
using the results by Haensel \& Pichon (1994) based on new
laboratory measurements of masses of neutron--rich nuclei.
At higher densities, in the inner NS crust,
we use the results of Negele \& Vautherin
(1973) derived on the basis of a modified Hartree-Fock method.
Small discontinuities of the curves
in Fig.\ 1 are due to
the jumps of the nuclear composition.
Notice, that the properties of the neutrino
synchrotron emission vary rather smoothly
in the transition regions from domain ${\cal A}$
to ${\cal B}$ and from ${\cal B}$ to ${\cal C}$.

The {\it high-temperature} domain ${\cal A}$ is defined as
$T_P \ll T \ll T_{\rm F}$; it is realized for not too
high densities and magnetic fields where $T_P \ll T_{\rm F}$.
In Fig.\ 1, this domain exists only for
$B=10^{12}$~G at $\rho < 10^{11}$~g~cm$^{-3}$.
In domain ${\cal A}$, the degenerate electrons emit neutrinos
through many cyclotron harmonics; typical harmonics is $ s \sim x^3$.
Corresponding neutrino energies
$\omega \sim \omega_B^\ast x^3 \ll T$ are not restricted
by the Pauli principle. The quasiclassical approach of KLY
yields
\begin{eqnarray}
   Q_{\rm syn}^{\rm A} & = & {2 \over 189 \pi^5} \;{ G_{\rm F}^2
               k_{\rm B} T m_e^2 \omega_B^6 x^8 \over c^5 \hbar^4} \,
               (25C_+^2-21C_-^2)
\nonumber \\
     & \approx & 3.09 \times 10^{15} \,
       B_{13}^6 T_9 x^8~~{\rm ergs~cm^{-3}~s^{-1}},
\label{QA}
\end{eqnarray}
where $T_9=T/(10^9~{\rm K})$. Here and in what follows,
numerical factors in the practical expressions are slightly
different from those presented by KLY because now we use more
accurate value of the Fermi constant (see Eq.\ (\ref{Quantum})).

The {\it moderate-temperature} domain ${\cal B}$ is defined as
$T_B \la T \ll T_P$ and $T \ll T_{\rm F}$.
It covers wide temperature and density ranges (Fig.\ 1)
most important for 
applications. 
In this domain, neutrinos are again emitted
through many cyclotron harmonics
$s \sim k_{\rm B} T/ \hbar \omega_B^\ast \gg 1$,
but their spectrum is restricted by the Pauli
principle, and typical neutrino energies are
$\omega \sim k_{\rm B} T$.
As shown by KLY, in this case the neutrino emissivity
is remarkably independent of the electron number density:
\begin{eqnarray}
   Q_{\rm syn}^{\rm B} & = & {2 \zeta(5) \over 9 \pi^5} \;{ G_{\rm F}^2
               m_e^2 \omega_B^2  \over c^5 \hbar^8 } \,
               C_+^2 \, (k_{\rm B} T)^5
\nonumber \\
     & \approx & 9.04 \times 10^{14} \,
       B_{13}^2 T_9^5~~{\rm ergs~cm^{-3}~s^{-1}},
\label{QB}
\end{eqnarray}
where $\zeta(5) \approx 1.037$ is the value of the Riemann
zeta function.

The third, {\it low-temperature} domain ${\cal C}$ corresponds
to temperatures $T \la T_B$ at which the main contribution
into the 
neutrino 
synchrotron emission comes from a few lower cyclotron
harmonics $s$=1, 2, \ldots ~~ If $T \ll T_B$, even the first harmonics
$s=1$ appears to be exponentially suppressed as discussed by
KLY. A more detailed analysis will be given below.

The emissivity $Q_{\rm syn}^{\rm AB}$ in the combined domain
${\cal A}+{\cal B}$, including a smooth transition from
${\cal A}$ to ${\cal B}$ at $T \sim T_P$, was calculated
accurately in KLY. The results (Eqs.\ (13), (15) and (18) in KLY),
valid at $T_B \la T \ll T_{\rm F}$,
can be conveniently rewritten as
\begin{eqnarray}
    Q_{\rm syn}^{\rm AB} & = & Q_{\rm syn}^{\rm B} S_{\rm AB},
\label{QAB} \\
    S_{\rm AB}  & = & {27 \xi^4 \over \pi^2 \, 2^9 \, \zeta(5)}
       \left[F_+(\xi) - {C_-^2 \over C_+^2} F_-(\xi) \right],
\nonumber     \\
    \xi & \equiv & {T_P \over T} = {3 \over 2} z x^3;  \hspace{5mm}
     z = {T_B \over T},
\label{SAB}
\end{eqnarray}
where the analytic fits to the functions $F_\pm(\xi)$ read
\begin{eqnarray}
  F_+(\xi) & = & D_1 \, { (1+c_1 y_1)^2 \over
        (1+ a_1 y_1 + b_1 y_1^2)^4},
\nonumber \\
            & \; \; &   \;  \;  \;
\nonumber  \\
  F_-(\xi) & = & D_2 \, { 1 + c_2 y_2 + d_2 y_2^2 + e_2 y_2^3
        \over (1+ a_2 y_2 + b_2 y_2^2)^5 }.
\label{F}
\end{eqnarray}
In this case
$y_{1,2} = [(1+ \alpha_{1,2} \xi^{2/3})^{2/3}-1]^{3/2}$,
$a_1=2.036 \times 10^{-4}$,
$b_1=7.405 \times 10^{-8}$,
$c_1=3.675 \times 10^{-4}$,
$a_2=3.356 \times 10^{-3}$,
$b_2=1.536 \times 10^{-5}$,
$c_2=1.436 \times 10^{-2}$,
$d_2=1.024 \times 10^{-5}$,
$e_2=7.647 \times 10^{-8}$,
$D_1=44.01$, $D_2=36.97$,
$\alpha_1 = 3172$, $\alpha_2 = 172.2$.

Now let us study the neutrino emissivity
in the combined domain ${\cal B}+{\cal C}$.
Using the quasiclassical expressions of KLY
in the ultrarelativistic limit ($m_e \to 0$) we obtain
\begin{equation}
    Q_{\rm syn}^{\rm BC} = Q_{\rm syn}^{\rm B} S_{\rm BC},
\label{QBC}
\end{equation}
where
\begin{eqnarray}
    S_{\rm BC}  & = & {3  \over  2^6  \zeta(5)} \,{1  \over  z^2 \, T^7}
         \sum_{s=1}^\infty
         \int_0^\pi \sin^3 \theta \, {\rm d}\theta
          \int q_\perp \, {\rm d}q_\perp
\nonumber    \\
     & \; \; & \times~\int {\rm d}q_z  (\omega^2 - q_z^2 - q_\perp^2)
         (\Psi - \Phi) { \omega^2 \over {\rm e}^{\omega/T}-1 }
\label{SBC}
\end{eqnarray}
with
\begin{equation}
     \Psi-\Phi  =  2 \left[ \left( {s^2 \over y^2} - 1 \right)
               J_s^2(y) +  J_s^{\prime 2}(y)  \right].
\label{J}
\end{equation}
Here, $J_s(y)$ is a Bessel function of argument
$y=p_\perp q_\perp c/(eB)$, and $J'_s(y)= {\rm d}J_s(y)/{\rm d}y$.
The neutrino-pair energy can be expressed as
$\omega \approx q_z \cos \theta + s \omega_B^\ast$.
The integration should be done over the
kinematically allowed domain:
$\kappa_\perp^2 \geq q_\perp^2 +
\sin^2 \theta (\kappa_z - q_z)^2$, $\kappa_\perp =
s \omega_B^\ast / \sin \theta$, $\kappa_z = s \omega_B^\ast \,
\cos \theta / \sin^2 \theta$.

One can easily see that $S_{\rm BC}$ depends on the
only parameter $z$. The
domain ${\cal B}$ corresponds to $ x^{-3} \ll z \ll 1$ while
the domain ${\cal C}$ corresponds to $z \ga 1$.
Equation (\ref{SBC}), in which we set $m_e \to 0$,
does not reproduce the high-temperature domain ${\cal A}$.
However, we have already
described the transition from ${\cal B}$ to ${\cal A}$ by
Eqs.\ (\ref{QAB}--\ref{F}).

At $z \ll 1$, 
the neutrino emissivity, determined by 
$S_{\rm BC}$ in  Eq.\ (\ref{SBC}), 
comes from many cyclotron harmonics (see KLY, for details).
A Bessel function $J_s(y)$ and its derivative
can be replaced by McDonald functions.
The main 
contribution to 
the integrals (\ref{SBC})
comes from a narrow vicinity $|\kappa_\perp - q_\perp|
\la \omega_B^\ast s^{1/3}$, and $|q_z - \kappa_z| \la
\omega_B^\ast s^{2/3}$ of the saddle point
$q_\perp = \kappa_\perp$ and $q_z = \kappa_z$, in which
the neutrino-pair energy is nearly constant,
$\omega \approx s \omega_B^\ast /\sin^2 \theta$.
Adopting these approximations and replacing
the sum over $s$ by the integral we reproduce evidently
the result of KLY, $S_{\rm BC}=1$. However if we take into account
small corrections in the expression of a Bessel
function through McDonald functions (Sokolov \&
Ternov 1974), and weak variation of $\omega$
near the saddle-point, we obtain
a more accurate asymptote
$ S_{\rm BC} = 1 - 0.4535 \, z^{2/3}$. The derivation is
outlined in Appendix~B.

In the opposite limit of $z \gg 1$, one can
keep the contribution from the first harmonics $s=1$, and
replace $(\Psi - \Phi) \to 1$,~~~
$({\rm e}^{\omega /T} -1)^{-1} \to {\rm e}^{- \omega / T}$
in Eq.\ (\ref{SBC}) as described in KLY.
At any kinematically allowed $q_z$ and $q_\perp$,
the neutrino-pair emission is
suppressed by a small factor $ \sim {\rm e}^{-\omega/T}$.
As shown in KLY, the most efficient
neutrino synchrotron radiation occurs from a small vicinity
of the allowed region, where
$q_z \approx -\omega^\ast_B/(1+\cos\theta)$ has minimum
and the neutrino energy
$\omega \approx \omega^\ast_B/(1+ \cos \theta)$ is
most strongly reduced by the quantum recoil effect.
This region corresponds to the backward electron scattering.
The integration over $q_\perp$ is then done
analytically, and we are left with a two-fold
integration over $q_z$ and $\theta$.
In the limit of very high $z$, it gives
\begin{equation}
           S_{\rm BC}
        \approx
           {3 \over 2 \zeta(5)} \,
              \exp \left( -{z \over 2} \right) \,
              \left( 1 + {28 \over z} \right).
\label{SCas}
\end{equation}
The convergence of this asymptote is very slow,
and we present the second-order correction term $28/z$ which
improves the convergence considerably. For instance, at $z=60$
the two-term asymptote gives an error of about 6.5\%,
while the one-term asymptote gives an error of about 36\%.

Notice, that the emissivity given by Eq.\ (20)
in KLY in the limit of $z  \gg 1$ is 4 times
smaller than the correct
emissivity presented here (due to a simple omission
in evaluating $Q_{\rm syn}^{\rm C}$  made by KLY).
Thus Eqs.\ (20) and (21) in KLY are inaccurate at $z \gg 1$.

In addition to analyzing the asymptotes, we have calculated
$S_{\rm BC}$ numerically from Eq.\ (\ref{SBC}) in the quasiclassical
approximation at intermediate $z$. The results are fitted by
the analytic expression
\begin{equation}
    S_{\rm BC}= \exp(-z/2)\, D_1(z)/D_2(z),
\label{Sfit}
\end{equation}
where $D_1(z)=1+0.4228 \,z +0.1014 \, z^2+ 0.006240 \, z^3$,
$D_2(z)=1+0.4535\, z^{2/3}+0.03008 \, z - 0.05043 \,
z^2 + 0.004314 \, z^3$.
The fit reproduces both the low-$z$ and the high-$z$ asymptotes.
The rms fit error at $z \leq 70$ is about 1.6\%, and
the maximum error is 5\% at $z \approx 18$.

Now, we can easily combine Eqs.\ (\ref{QAB}) and
(\ref{QBC}) and obtain a general fit expression
for the neutrino synchrotron emissivity
which is valid everywhere
in domains ${\cal A}$, ${\cal B}$, ${\cal C}$ ($T \ll T_{\rm F}$,
$ \rho \gg \rho_B$),
where the electrons are degenerate,
relativistic and populate many Landau levels:
\begin{equation}
   Q_{\rm syn}^{\rm ABC} = Q_{\rm syn}^{\rm B} \,
                           S_{\rm AB} \, S_{\rm BC}.
\label{ABCfit}
\end{equation}
Here, $Q_{\rm syn}^{\rm B}$ is given by Eq.\ (\ref{QB}),
while $S_{\rm AB}$ and $S_{\rm BC}$ are defined by Eqs.\ (\ref{SAB}),
(\ref{F}) and (\ref{Sfit}).

\begin{figure}
\vskip 0.8cm
\hskip -1cm
\epsfxsize=8.0cm
\epsfbox{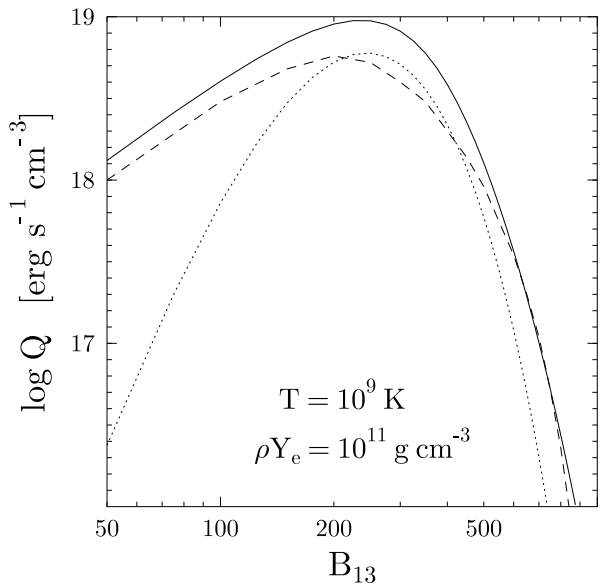}
\vskip -1cm
\caption[ ]{
Neutrino synchrotron emissivity vs $B$ from a plasma
with $\rho \, Y_e = 10^{11}$~g~cm$^{-3}$ and
$T=10^9$~K. Solid curve shows our quasiclassical
result, dash--curve
--- calculation of Vidaurre et al.\ (1995),
dots --- analytic Eq.\ (26) of these authors.
}
\label{fig2}
\end{figure}

\begin{figure}
\hskip -0.5cm
\epsfxsize=9.0cm
\epsfbox{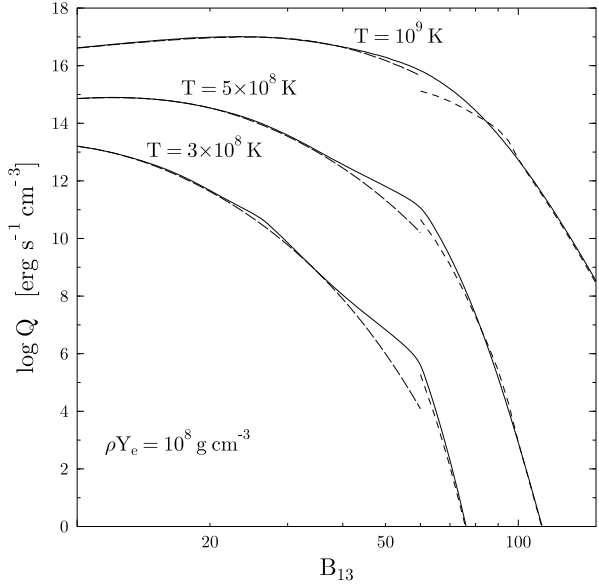}
\vskip -1cm
\caption[ ]{
Neutrino synchrotron emissivity versus $B$ from a plasma
with $\rho \, Y_e = 10^8$~g~cm$^{-3}$ at
$T=3 \times 10^8$, $5 \times 10^8$ and $10^9$~K. Solid curves are
calculated with the quantum code, long--dash--curves 
are obtained using the 
quasiclassical calculation valid at $B \ll 6 \times 10^{14}$~G,
dash--curves show the quantum asymptotic expression (Appendix~C)
for the case in which the ground Landau level is populated alone, 
$B > 6 \times 10^{14}$~G.
}
\label{figQ}
\end{figure}

\begin{figure}
\hskip -0.5cm
\epsfxsize=8.0cm
\epsfbox{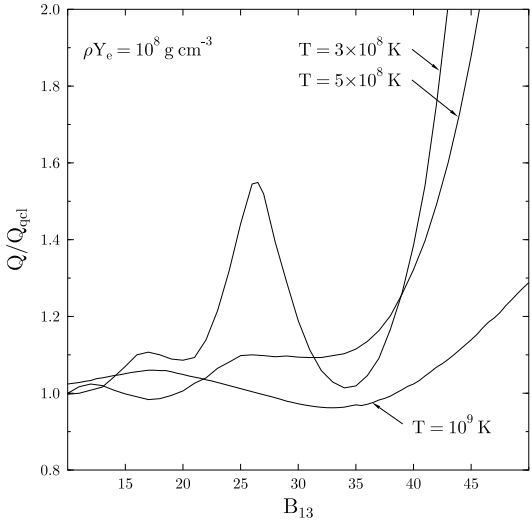}
\vskip -1cm
\caption[ ]{
Ratios of the quantum to quasiclassical neutrino synchrotron
emissivities presented in Fig. \protect{\ref{figQ}}.
}
\label{figR}
\end{figure}

The neutrino synchrotron emission
in domains ${\cal B}+{\cal C}$ (in our notations),
has been reconsidered recently
by Vidaurre et al.\ (1995).
Their results are compared with ours 
in Fig.\ 2 which
shows the 
neutrino synchrotron 
emissivity as a function of
$B$ for the same plasma parameters ($\rho Y_e
= 10^{11}~{\rm g~cm}^{-3}$, $T=10^9$~K) as in Fig.\ 4
by Vidaurre et al.\ (1995). For the parameters chosen
(due to very high $\rho$),
domain ${\cal A}$ is not realized, while
the transition from domain
${\cal B}$ to ${\cal C}$ occurs at fairly
high $B$ (for instance, $z=1$ corresponds to
$B \approx 3.5 \times 10^{14}$~G).
Our quasiclassical calculation, analytic fit
(\ref{ABCfit}) and quantum calculation are
so close that yield the same (solid) line.
Numerical calculation of Vidaurre et al.\ (1995)
is shown by dashes, and their analytical
approximation (claimed to be accurate at
intermediate $z$) by dots. In our notations,
the latter approximation reads $S_{\rm BC}^{\rm (V)}
=0.0073 \, z^5 \, {\rm e}^{-z}$. 
We see that 
such an approximation 
reproduces neither low--$z$ nor high--$z$
asymptote, and is, in fact, inaccurate at
intermediate $z$. It 
also 
disagrees with numerical curve of
Vidaurre et al.\ (1995) and with our curve.
The numerical results by Vidaurre et al.\ (1995) are 
considerably different from ours except at
$B \approx (6 - 8) \times 10^{15}$~G.
At lower and higher $B$
the disagreement becomes substantial.
In domain ${\cal B}$, Vidaurre et al.\ (1995)
present another analytic expression that differs
from Eq.\ (\ref{QB}) by a numerical factor $9/(2 \pi)$. The difference
comes from two inaccuracies made by Vidaurre et al.\ (1995).
First, while deriving the emissivity,
they  use the asymptote of the McDonald function
at large argument (their Eq.~(19)) 
instead of exact expression for the McDonald function
(as was done by KLY).
One can easily verify that this inaccuracy yields an extra factor
$3 / \pi$.
Secondly, Vidaurre et al.\ (1995) calculated inaccurately 
an integral over $q_z$ (their
Eq.\ (22)) as if the integrand were independent of $q_z$.
This yields the second extra factor 3/2.
Thus we 
can conclude that the results by Vidaurre et al.\ (1995)
are rather inaccurate in a wide parameter range.
We have performed extensive comparison
of the results obtained with our quantum code and with the
quasiclassical approach. We have found very good agreement
(within 3--5 \%) in all the cases in which the quasiclassical
approach can be used (see above). This statement is illustrated
in Figs.\ \ref{figQ} and \ref{figR}. Figure \ref{figQ}
shows the quantum and quasiclassical synchrotron emissivities
from a plasma with $\rho \, Y_e = 10^8$~g~cm$^{-3}$
and $T=10^9$~K as a function of $B$. Figure \ref{figR}
displays the ratio of the quantum to quasiclassical emissivities
in more detail. The quasiclassical
approach is valid as long as $\rho \gg \rho_B$
(as long as electrons populate many Landau levels)
which corresponds
to $B \ll 6 \times 10^{14}$~G for our particular parameters.
For these magnetic fields,
the quantum and quasiclassical results are seen to coincide
quite well. Low-amplitude oscillations of the curves
in Fig.\ \ref{figR} reflect oscillations
of the synchrotron emissivity as calculated with the quantum code.
The oscillations are produced by depopulation of
higher Landau levels with increasing $B$. They represent
quantum effect associated with square--root singularities
of the Landau states. As seen from Fig.\ \ref{figR}
the oscillations are smeared out
with increasing $T$ by thermal
broadening of the square-root singularities (
cf. 
Yakovlev \& Kaminker 1994).
In a higher field $B \ga 6 \times 10^{14}$~G, the electrons
populate the ground Landau level alone and
the electron chemical potential is reduced by the magnetic field.
Very high $B$, not shown in Fig.\ \ref{figQ},
remove the electron degeneracy.
Note that at 
high temperatures, one should take into account
the synchrotron neutrino emission by positrons
(Kaminker \& Yakovlev 1994).
All these conditions are described by the quantum code while
the quasiclassical approach is no longer valid.
In Appendix~C, we obtain simple asymptotic expressions
for $Q_{\rm syn}$ in the case in which the electrons populate the ground
Landau level alone. Corresponding curves are shown by
dashed curves in Fig.\ \ref{figQ}, and they
are seen to reproduce the
exact quantum curves quite accurately.

\section{Discussion and results}
%
\begin{figure}
\vskip 0.8cm
\hskip -1cm
\epsfxsize=8.0cm
\epsfbox{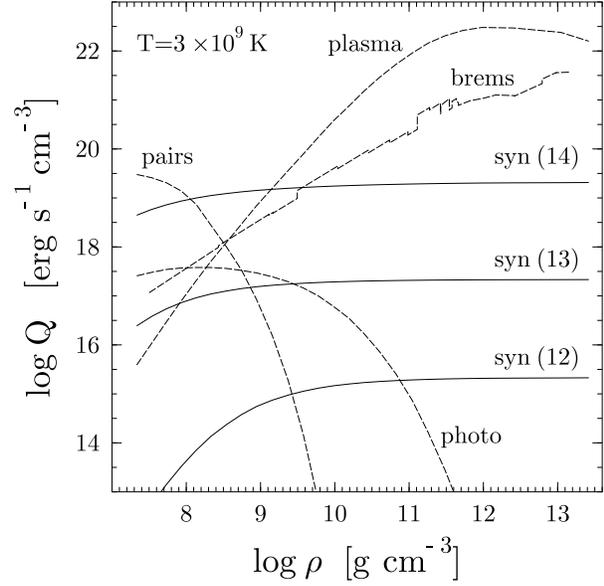}
\vskip -1cm
\caption[ ]{
Density dependence of neutrino emissivities
from ground-state matter of the NS crust
due to various mechanisms at $T= 3 \times 10^9$~K.
Curves `syn' `(12)', `(13)', and `(14)'
refer to the synchrotron mechanism at
$B=$ 10$^{12}$, 10$^{13}$ and 10$^{14}$~G, respectively.
Curve `pairs' corresponds to neutrino emission due
to annihilation of electron-positron pairs; it is almost
independent of $B$ at given $T$. Other curves are for $B=0$:
`brems' --- the total electron--nucleus bremsstrahlung;
`plasma' --- plasmon decay;
`photo' --- photoneutrino process (see text).
}
\label{fig93}
\end{figure}
%

\begin{figure}
\vskip 0.8cm
\hskip -1cm
\epsfxsize=8.0cm
\epsfbox{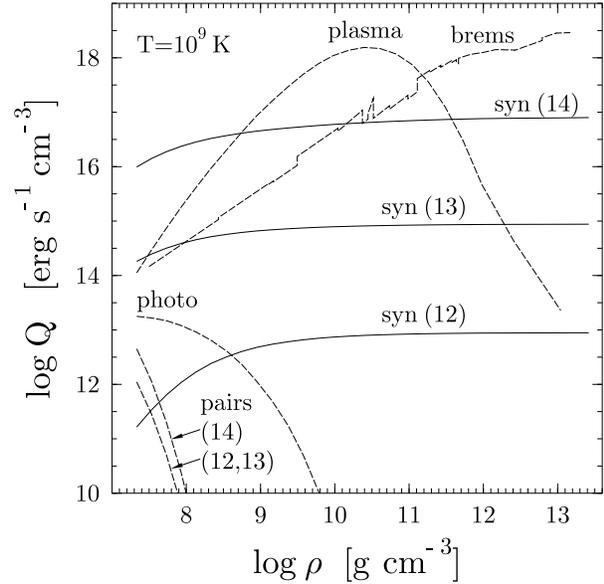}
\vskip -1cm
\caption[ ]{
Same as in Fig.\ \protect{\ref{fig93}} but
at $T= 10^9$~K. Pair annihilation depends noticeably on $B$.
}
\label{fig91}
\end{figure}
%

\begin{figure}
\vskip 0.8cm
\hskip -1cm
\epsfxsize=8.0cm
\epsfbox{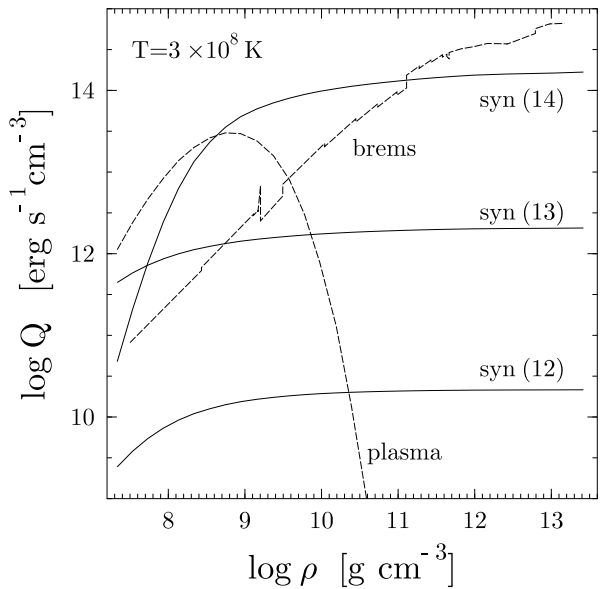}
\vskip -1cm
\caption[ ]{
Same as in Figs.\ \protect{\ref{fig93}} and \protect{\ref{fig91}}
but at $T= 3 \times 10^8$~K. Pair annihilation and photoneutrino
processes become negligible.
}
\label{fig83}
\end{figure}
%

\begin{figure}
\vskip 0.8cm
\hskip -1cm
\epsfxsize=8.0cm
\epsfbox{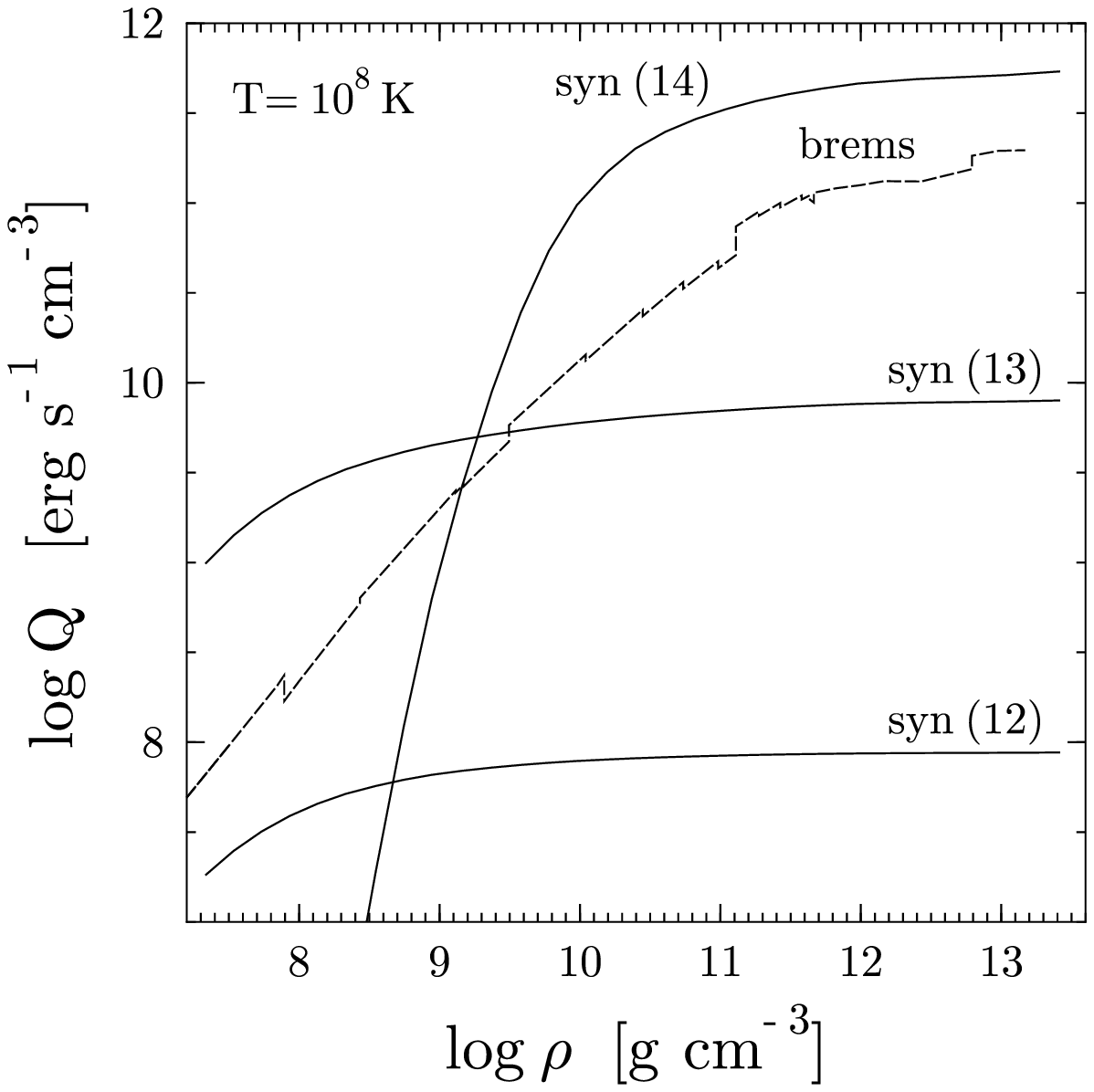}
\vskip -1cm
\caption[ ]{
Same as in Figs.\ \protect{\ref{fig93}} -- \protect{\ref{fig83}}
but at $T= 10^8$~K. Plasmon decay becomes negligible.
}
\label{fig81}
\end{figure}
%

Figures \ref{fig93}--\ref{fig81} display
density dependence of the neutrino synchrotron
emissivity $Q_{\rm syn}$ calculated from Eq.\ (\ref{ABCfit})
for the magnetic fields $B=10^{12}$, 10$^{13}$, and 10$^{14}$~G
at four temperatures
$T=3 \times 10^9$, $T=10^9$, $3 \times 10^8$,
and  $10^8$~K, respectively.
We adopt the ground--state model of matter in the NS
crust (see Sect.\ 3). Various neutrino--emission regimes
can be understood by comparison with Fig.\ 1.

The high-density (horizontal) parts
of the synchrotron curves
correspond to domain ${\cal B}$ (Eq.\ (\ref{QB})),
where $Q_{\rm syn}$ is density independent.
The low-density bends
are associated with transitions either into
domain ${\cal A}$ (where $Q_{\rm syn} \propto \rho^{8/3}$
according to Eq.\ (\ref{QA})) or into domain ${\cal C}$ (where
$Q_{\rm syn}$ decreases exponentially due to
cyclotron harmonics suppression, Eq.\ (\ref{SCas})).
Domain ${\cal A}$ is realized only
for $B = 10^{12}$~G and $T \ga 10^9$~K
in Figs.\ \ref{fig93} and \ref{fig91}. 
The low-density bends of $Q_{\rm syn}$
in domain ${\cal C}$ are 
much steeper than those
in domain ${\cal A}$.
%
These bends are more 
pronounced at highest $B=10^{14}$~G, 
at which domain ${\cal C}$ extends to higher $T$ and $\rho$
(Figs.\ \ref{fig83} and \ref{fig81}).

For comparison, we also plot the emissivities produced by
other neutrino generation mechanisms: the electron-positron
pair annihilation
into neutrino pairs,
electron--nucleus bremsstrahlung, plasmon decay
and photon decay.
The pair annihilation in a magnetized plasma
has been considered by Kaminker et al.\ (1992a, b),
and Kaminker \& Yakovlev (1994).
For the parameters of study,
the emissivity appears to be 
weakly 
dependent on the magnetic field. At $B \la 10^{13}$ G it is very
close to the zero--field emissivity (Itoh et al.\ 1989, 1996).
As seen from Figs.\
\ref{fig93} and \ref{fig91}, the pair--annihilation emissivity
differs slightly from the zero-field one only
in a not too hot plasma ($T \la 10^9$~K) at $B \ga 10^{14}$~G.
The
neutrino
bremsstrahlung curves are plotted neglecting the
influence of the magnetic field. The effect of the
field on the
bremsstrahlung has not been
studied so far but it is expected to be
weak, for the parameters in Figs.\ \ref{fig93}--\ref{fig81}.
We use the results of Haensel et al.\ (1996) to describe
the neutrino pair bremsstrahlung due to Coulomb scattering
of electrons by atomic nuclei in the liquid phase of matter.
In the solid phase, similar process is known to consist
of two parts: the phonon and static lattice contributions.
We use the results by Yakovlev \& Kaminker (1996) to evaluate the
phonon contribution. As for the static lattice contribution,
we employ the most recent theory by Pethick \& Thorsson (1996)
and perform numerical calculation
from Eqs.\ (28) and (29) of their paper (adopting
the Debye--Waller factor and
the nuclear form-factor which were used
by Yakovlev \& Kaminker 1996).
Numerous jumps of the bremsstrahlung
curves in Figs.\ \ref{fig93}--\ref{fig81} are associated
either with jump-like changes of nuclear composition
of cold--catalyzed matter or with solid--liquid phase
transitions (see Haensel et al.\ 1996 for details).
The neutrino emissivities from other processes are
determined by the electron and positron number densities which are
nearly continuous function of the density. Therefore,
all other curves are smooth.
The neutrino generation
due to plasmon and photon decays in a magnetic field
has not been considered in the literature, and we
present the field--free results of Itoh et al.\ (1989, 1996),
for illustration.

In the case of zero magnetic field,
the bremsstrahlung process dominates completely
in most dense layers of the NS crust
at not too high temperatures 
$T \la 10^9$~K.
Plasmon decay,
photon decay, and pair annihilation
are significant at high temperatures,
$T \ga 10^9$~K, but their emissivities become negligible
very soon as temperature decreases.

The synchrotron emissivity is, to some extent, similar
to the bremsstrahlung, for it persists over the wide temperature
and density ranges.
In the presence of the strong magnetic field $B \ga 10^{13}$~G,
the synchrotron emission is seen to be
important and even dominant for any $T$ in Figs.\
\ref{fig93}--\ref{fig81}. In a hot plasma (Fig.\ \ref{fig93}),
the synchrotron emission is significant at comparatively
low densities, $\rho = 10^8$--$10^9$~g~cm$^{-3}$.
With decreasing $T$ the 
neutrino synchrotron emission
becomes more important at higher densities.
At $T=10^8$~K, only the bremsstrahlung and
synchrotron emissions actually survive (Fig.\ \ref{fig81});
if $B=10^{14}$~G, 
the synchrotron emission 
dominates over the bremsstrahlung one in a wide
density range, $\rho \ga 10^9$~g~cm$^{-3}$.

%
\section{Conclusions}
We have considered the synchrotron neutrino-pair emission
of electrons from a dense magnetized plasma.
We developed a computer code (Sect.\ 2) which calculates the
synchrotron emission 
in the presence of
quantizing magnetic fields 
for a wide range of conditions at which the electrons can be weakly 
as well as strongly degenerate and/or relativistic. 
We have also calculated the
synchrotron emissivity in the quasiclassical approximation
(Sect.\ 3) for strongly degenerate, relativistic electrons
which populate many Landau levels.
We have paid special attention to the case in which 
the main contribution into the synchrotron neutrino emission comes 
from the fundamental 
or several 
low 
cyclotron harmonics. 
This case has not been analyzed
properly earlier by KLY and Vidaurre et al.\ (1995).
We have obtained a simple
analytic expression (\ref{ABCfit}) which fits accurately
our quasiclassical results in wide ranges of densities,
temperatures, and magnetic fields.
In Sect.\ 4 we have demonstrated that the synchrotron
neutrino emissivity gives considerable or even dominant
contribution into the neutrino emissivity
at $B \ga 10^{13}$~G in moderate-density and/or
high-density layers of the NS crusts, depending
on temperature and magnetic field.
Let us notice, that the neutrino emissivities plotted
in Figs.\ 5--8 are appropriate for cooling neutron
stars rather than for newly born or merging
neutron stars. Since 
the neutrino 
radiation
from a cooling neutron star 
is too 
weak 
to be detected
with modern neutrino observatories
we do not calculate the spectrum of emitted neutrinos.

Our results indicate that
the neutrino synchrotron emission
should be taken 
into account in the cooling theories of magnetized
neutron stars. It can be important
during initial cooling phase ($t$=
10--1000
~yrs after the
neutron star birth). At this stage, the thermal
relaxation of internal stellar layers is not achieved
(e.g., Nomoto \& Tsuruta 1987, Lattimer et al.\ 1994)
and local emissivity
from different crustal layers can affect this relaxation
and observable surface temperature.

The synchrotron
process in the crust can also be significant at
the late neutrino cooling stages ($t \sim 10^5$~yrs,
$T \sim 10^8$~K),
in the presence of strong magnetic fields
$B \ga 10^{14}$~G. The synchrotron emission
can be the dominant neutrino
production mechanism in the neutron star crust,
while the neutrino luminosity from the stellar
core may be not too high, at this cooling stage,
and quite comparable with the luminosity from the crust.

It is worthwhile to mention
that the synchrotron emission
can be important also in the superfluid neutron star cores
(Kaminker et al.\ 1997). A strong superfluidity of
neutrons and protons suppresses greatly
(e.g., Yakovlev \& Levenfish 1995) the traditional
neutrino production mechanisms
such as Urca processes or nucleon--nucleon bremsstrahlung.
The superfluidity (superconductivity) of protons
splits an initial, locally uniform magnetic field
of the neutron star core into fluxoids --- thin
magnetic threads of quantized magnetic flux.
This process modifies the neutrino synchrotron process
(Kaminker et al.\ 1997) amplifying it
just after the superconductivity onset and making it
significant since the traditional neutrino
generation mechanisms are suppressed.
It has been shown that this modified neutrino synchrotron process
can dominate over other mechanisms if the
initially uniform magnetic
field in the NSs core is $B \ga 10^{13}$~G and
if $T \la 5 \times 10^8$~K.
These fields and temperatures are quite consistent with
the results of the present article:
similar conditions are necessary for
the synchrotron emission to dominate over other
neutrino production mechanisms in the NS crust by the
end of the neutrino cooling stage. If so,
the total neutrino luminosity (from the stellar crust and core)
can be governed by internal stellar magnetic fields
which can affect the neutron star cooling.
We plan to study this cooling in a future article.

{\bf Acknowledgements}
We are grateful to A.Y.\ Potekhin for providing us with
the program which calculates the electron chemical potential.
Two of the authors (VGB and ADK) acknowledge excellent working
conditions and hospitality of N.\ Copernicus Astronomical Center in Warsaw.
This work was supported in part by  KBN (grant 2P 304 014 07),
RBRF (grant No.\ 96-02-16870a), INTAS (grant No.\ 94-3834), and
DFG-RBRF (grant No.\ 96-02-00177G).

\renewcommand{\theequation}{A\arabic{equation}}
\setcounter{equation}{0}
\section*{Appendix A}

For performing exact quantum calculation
of the neutrino synchrotron emissivity from Eq.\ (\ref{Quantum})
we replace the integrations over
$p_z$ and $q_z$ by the integrations over
initial and final electron energies $\varepsilon$ and $\varepsilon'$,
respectively. In this way we can accurately integrate
within the intervals 
$|\varepsilon- \mu| \la T$ and $|\varepsilon'- \mu| \la T$, 
where the Fermi-Dirac distributions are 
rapidly varying. 
It is also convenient to
set $q_\perp^2 = \left( \omega^2 - q_z^2 \right)(1- \eta)$ and replace
the integration over $q_\perp$ by the integration over $\eta$.
Then
\begin{eqnarray}
        Q_{\rm syn}  &=& \frac{G_{\rm F}^2 b}{6 (2\pi)^5}
        \sum_{n=1}^\infty
        \int_{\varepsilon_n}^\infty
        \frac{{\rm d} \varepsilon}{\sqrt{\varepsilon^2
        - \varepsilon_n^2}} \, f
\nonumber    \\
       &\times&    \sum_{s=1}^n
        \int
        \frac{{\rm d} \varepsilon'}{\sqrt{\varepsilon'^2
        - \varepsilon_{n'}^2}} \,
        \omega \left( \omega^2 - q_z^2 \right)^2 D (1 - f')  ,
\nonumber \\
        D & = & \int_0^1 {\rm d} \eta
        \left( C_+^2 R_+ - C_-^2 R_- \right) ,
\nonumber \\
        R_+ & = &
        \left\{ 1 + \eta \left[ 1 + p_\perp^2 + p_\perp^{\prime 2} -
        \left( \omega^2 -q_z^2 \right) \eta
        \right] \right\} \Psi
\nonumber \\
         & - &
        \left[ 1 + \left( p_\perp^2 + p_\perp^{\prime 2} \right) \eta
        \right] \Phi ,
\nonumber \\
          & \; \; & \; \; \;
\nonumber   \\
        R_- & = & (1+\eta) \Psi - (1-2\eta) \Phi .
\label{Q_general}
\end{eqnarray}
Here, $n'=n-s$ ($s$ enumerates cyclotron harmonics);
$\varepsilon_n=\sqrt{1+2nb} \,$ and
$\varepsilon_{n'}=\sqrt{1+2n'b} \,$ correspond to the
excitation (de-excitation) thresholds of the initial and final
electron states, respectively.
The neutrino pair energy and longitudinal momenta are determined from
conservation laws:
$p_z=\sqrt{\varepsilon^2-\varepsilon_n^2} \,$ and
$p'_z=\pm \sqrt{\varepsilon^{\prime 2}-\varepsilon_{n'}^2} \,$.
The sign of $p'_z$ and
the domain of integration over $\varepsilon'$ are
specified by the
minimum longitudinal momentum of the final-state electron
for given $n$, $\varepsilon$ and $s$:
\begin{equation}
        p'_1  \leq p'_z  \leq p'_2 , \; \; \;
        p'_{1,2} = \pm
       \frac{\varepsilon_{n'}^2 - (\varepsilon \mp p_z)^2}
       {2(\varepsilon \mp p_z)}
\end{equation}

If $p'_1 \geq 0$, one has
$p'_z=\sqrt{\varepsilon^{\prime 2}-\varepsilon_{n'}^2} \,$ and
$\varepsilon'_1 \leq \varepsilon'\leq \varepsilon'_2$, where
\begin{equation}
      \varepsilon'_{1,2} =
      \frac{\varepsilon_{n'}^2 + (\varepsilon \mp p_z)^2}
      {2(\varepsilon\mp p_z)}.
\end{equation}
If $p'_1 < 0$, then there are two integration domains.
The first one is $\varepsilon_{n'} \leq \varepsilon' \leq \varepsilon'_1$,
with
$p'_z=-\sqrt{\varepsilon^{\prime 2}-\varepsilon_{n'}^2} \,$.
The second domain
is $\varepsilon_{n'} \leq \varepsilon' \leq \varepsilon'_2$, with
$p'_z=\sqrt{\varepsilon^{\prime 2}-\varepsilon_{n'}^2}$.

\renewcommand{\theequation}{B\arabic{equation}}
\setcounter{equation}{0}
\section*{Appendix B}

Consider the $z \to 0$ asymptote of the function $S_{\rm BC}(z)$
given by Eq.\ (\ref{SBC}). The integrand of (\ref{SBC})
contains the factor $R=\omega^2/(\exp(\omega/T)-1)$.
Since the neutrino-pair energy $\omega=q_z \cos \theta + s \omega_B^\ast$
varies near the saddle point, we have
\begin{equation}
   R(\omega)=R(\omega_0) + R'\, \Delta\omega
            + {1 \over 2} R'' \, \Delta\omega^2,
\label{f}
\end{equation}
where
$\omega_0=\kappa_z+s \omega_B^\ast$ is the
value of $\omega$ in the saddle point,
$\Delta\omega= \cos \theta (q_z - \kappa_z)$,
\begin{equation}
   R'' = {2+{\rm e}^v (-4+4v+v^2)+{\rm e}^{2v}(2-4v+v^2)
          \over ({\rm e}^v -1)^3},
\label{R}
\end{equation}
$v=\omega_0/T$, and
$\kappa_z= s \omega_B^\ast \cos \theta / \sin^2 \theta$
is the $z$-coordinate of the saddle point. The term in (\ref{f}),
which is linear in $\Delta \omega$, does not 
contribute to 
(\ref{SBC}) due to integration over $q_z$.

In the low-$z$ 
limit (i.e., for $T\gg T_B$), 
the Bessel functions $J_s(x)$ and $J'_s(x)$
in (\ref{J}) can be expressed through McDonald functions
$K_{1/3}(\eta)$
and $K_{2/3}(\eta)$ of the argument
$\eta=(s/3)\epsilon^{3/2}$,
where $\epsilon=1-(x/s)^2=1-(q_\perp/\kappa_\perp)^2$;
$\kappa_\perp=\omega_B^\ast s /\sin \theta$ is the
coordinate of the saddle
point transverse to the magnetic field.
These expressions can be written as
(Sokolov \& Ternov 1974)
\begin{eqnarray}
   J_s(x) & = & { \sqrt{\epsilon} \over \pi \sqrt{3}}
      \left[K_{1/3} + {\epsilon \over 10}
      \left(K_{1/3} - 6 \eta K_{2/3} \right) \right],
\nonumber \\
   J'_s(x) & = & { \epsilon \over \pi \sqrt{3}}
      \left[K_{2/3} \right.
\nonumber \\
     &  + & \left. {\epsilon \over 5}
      \left(2 K_{2/3} -
      \left( {1 \over 3 \eta} + 3 \eta \right) K_{1/3} \right) \right].
\label{J1}
\end{eqnarray}
KLY took into account the main terms of these asymptotes.
Here, 
we include small corrections (proportional to
the factor $\epsilon \la s^{-2/3} \ll 1$
in square brackets). This yields
\begin{eqnarray}
    \Psi - \Phi  & = & {2 \epsilon^2 \over 3 \pi^2} \left[K_{1/3}^2+K_{2/3}^2
  +  { \epsilon \over 5} \left( 6 K_{1/3}^2 +  4 K_{2/3}^2
      \right. \right.
\nonumber \\
&-&  \left. \left.   2
      K_{1/3}K_{2/3} \left({1 \over 3 \eta} + 6 \eta
      \right) \right) \right].
\label{MD1}
\end{eqnarray}

Let us substitute (\ref{f}) and (\ref{MD1}) into (\ref{SBC}).
Then
\begin{equation}
    S_{\rm BC}(z) = 1 + a_1 z^{2/3} + a_2 z^{2/3} = 1 - 0.4535 z^{2/3}.
\end{equation}
In this case,
$ a_1 = L_1 M_1 N_1 /( 10 \, I_0)= 0.2053$
comes from variation of $\omega$, with
\begin{eqnarray}
   L_1 & = & \int_0^\pi {\rm d}\theta \, \sin^{5/3} \theta \, \cos^2 \theta \,
       = 0.45890,
\nonumber  \\
   M_1 & = & 3^{2/3} \int_0^\infty {\rm d}v \,
   R'' v^{10/3} = 3^{2/3} \times 76.533,
\nonumber \\
   N_1 & = & \int_0^\infty {\rm d}\eta \, \eta^{8/3} \, (K_{1/3}^2+K_{2/3}^2)=
   0.76706,
\nonumber \\
   I_0 & = & {8 \over 3} \pi^2 \zeta(5).
\label{a1}
\end{eqnarray}
Furthermore,
$ a_2 = L_2 M_2 N_2 /( 5 \, I_0) =-0.6588$
comes from the corrections to the McDonald functions,
\begin{eqnarray}
   L_2  & = & \int_0^\pi {\rm d} \theta \, \sin^{5/3} \theta = 1.6826,
\nonumber  \\
   M_2 & = & 3^{2/3} \int_0^\infty {\rm d}v \,
   {v^{10/3} \over {\rm e}^v -1} = 20.468,
\nonumber \\
   N_2 & = & \int_0^\infty {\rm d}\eta \, \eta^{8/3} \,
      \left[ 6 K_{1/3}^2  +   4  K_{2/3}^2 \right.
\nonumber \\
       & - &  \left. 2  K_{1/3}K_{2/3}
            \left({1 \over 3 \eta} + 6 \eta
           \right) \right]  = - 2.6082.
\label{a2}
\end{eqnarray}
%

\renewcommand{\theequation}{C\arabic{equation}}
\setcounter{equation}{0}
\section*{Appendix C}

Consider the neutrino synchrotron radiation by relativistic
degenerate electrons in a very strong magnetic field
($b \gg 1$, $\sqrt{1+2b}> \mu$)
in which 
the bulk of electrons 
populate the ground Landau level.
The electron Fermi momentum is
then given by $p_{\rm F} = 2x^3/(3b)$, and the chemical
potential is $\mu=\sqrt{1+p_{\rm F}^2}$. The number of electrons
on the excited Landau levels is exponentially small, and
the major 
contribution to 
the neutrino emission comes
from the electron transitions from the first
excited to the ground Landau level. Equation (\ref{A_general})
reduces to
\begin{equation}
    A = {C_+^2 \over \varepsilon \varepsilon'} \,
    [(1+b-u_0 b)(2bu_0-2bu+u)-u]\, {\rm e}^{-u},
\label{AA}
\end{equation}
where $u_0=(\omega^2 - q_z^2)/(2b)$. Inserting (\ref{AA}) into
(\ref{Quantum}) we can integrate over $q_{\perp}=\sqrt{2bu}$:
\begin{eqnarray}
   Q_{\rm syn} & = & {b^2 C_+^2 \over 3 (2 \pi)^5} \, G_{\rm F}^2 \,
   \int_{-\infty}^{+\infty} {\rm d}p_z \int_{\{u_0>0\}} {\rm d}q_z \,
   \omega f(1-f')
\nonumber \\
  & \times & { 1 \over \varepsilon \varepsilon'}
   \{(1+b-bu_0)[2bu_0+1-2b
\nonumber \\
  & + & (2b-1-u_0){\rm e}^{-u_0}]
   +  (1+u_0){\rm e}^{-u_0}-1 \}.
\label{QQ}
\end{eqnarray}
Since the first Landau level is almost empty, we can
set $f \approx \exp(-(\varepsilon - \mu)/t)$, where
$t=T/(m_e c^2)=T/(5.93 \times 10^9$ K)).
The neutrino emission
is greatly suppressed by the combination of Fermi-Dirac
distributions $f(1-f')$. These distributions determine
a very narrow integration domain which 
contributes to 
$Q_{\rm syn}$.

Equation (\ref{QQ}) can be 
further simplified 
in the two limiting cases.
The first case is $\mu_c < \mu < \sqrt{1+2b}$, or
$p_{\rm F}/2 < b < b_c$, where $\mu_c=(1+b)/\sqrt{1+2b}$ and
$b_c = p_{\rm F}(p_{\rm F}+\mu)$.
The main contribution into $Q_{\rm syn}$ comes from 
narrow vicinities 
of two equivalent saddle points
$p_{z0} = \pm [(p_{\rm F}+\mu)^2 - 2b-1]/[2(p_{\rm F}+ \mu)]$,
$q_{z0} = \mp b/(p_{\rm F}+\mu)$. Each point corresponds to
the most efficient electron transition in which
an initial-state electron descends to the ground Landau level
just with the Fermi energy $\varepsilon'_0= \mu$, 
emitting a neutrino-pair with the energy
$\omega_0=|q_{z0}|$. In this case,
the energy of the initial-state
electron is $\varepsilon_0 = \mu + \omega_0$.
One has $u_0 \ll 1$ in the
vicinities of the saddle points. Expanding $\varepsilon$,
$\varepsilon'$, and $u_0$ in these vicinities in powers
of $(p_z-p_{z0})$ and $(q_z-q_{z0})$, 
we obtain from Eq.\ (\ref{QQ})
\begin{equation}
   Q_{\rm syn} =
   { G_{\rm F}^2 \, C_+^2 b^4
   (2b+3) t^4 \over 3 (2 \pi)^4 p_{\rm F}^4 (p_{\rm F} +\mu) H^3 \, S}
         \exp \left( -{ b \over (p_{\rm F} + \mu) t} \right),
\label{high_mu}
\end{equation}
where
$H=1-(b/b_c)$ and $S=\sin( \pi H)$.

The second case corresponds to $\mu < \mu_c$ or $b>b_c$.
Now the most efficient electron transitions are those
in which the initial--state electron is near the bottom
of the first Landau level
($\varepsilon \approx \sqrt{1+2b}$, $p_z^2 \la t\sqrt{1+2b}$).
Accordingly, $\omega = \sqrt{1+2b}-\varepsilon'$ and
$u_0=(1+b-\varepsilon' \sqrt{1+2b})/b$.
The energy of the final--state electron $\varepsilon' \approx |q_z|$
varies mostly from the maximum energy
$\varepsilon'=(1+b)/\sqrt{1+2b}$
(associated with the minimum allowable
neutrino-pair energy $\omega=b/ \sqrt{1+2b}$ at $u_0=0$)
to the minimum energy
$\varepsilon'=\mu$ allowed by the Pauli principle.
One can put $p_z=0$ in all smooth functions under the integral
(\ref{QQ}), and replace integration over $q_z$ by integration
over $\varepsilon'$ or over $u$, with $(1-f')=1$ in the integration domain.
The integration is then taken analytically (for $b \gg 1$) yielding
\begin{eqnarray}
    Q_{\rm syn} & = &
    {G_{\rm F}^2 C_+^2  4 b^5 \sqrt{2 \pi t} \over 3 (2 \pi)^5 (1+2b)^{3/4}}
\nonumber \\
    & \times & F(u_1) \, \exp\left( - {\sqrt{1+2b}-\mu \over t}
              \right),
\label{low_mu}
\end{eqnarray}
where
\begin{equation}
    F(u_1)= {u_1^3 \over 3} - u_1 +2 - 2 \, {\rm e}^{-u_1} - u_1 \,
            {\rm e}^{-u_1},
\label{Fu}
\end{equation}
and $u_1 = 1-(\mu/\mu_c)$. 
In the limit of $\mu \to  1$ we have $u_1 \to 1$ and
  $F=(4/3)-(3/{\rm e}))$. 
Then Eq.\ (\ref{low_mu})
reproduces the asymptotic expression for the
synchrotron neutrino emissivity of electrons
from a non-degenerate electron-positron
plasma (Eq.\ (35) of the paper by Kaminker \& Yakovlev 1993).
In the opposite limit, $\mu \to \mu_c$, we obtain
$u_1 \to 0$ and $F \approx u_1^3/6$.

Both asymptotes, (\ref{high_mu}) and (\ref{low_mu}), become
invalid in a narrow vicinity $|1-(\mu/\mu_c)|
\approx 0.5|1-(b/b_c)| \la
\sqrt{t}/b^{1/4}$ of the point $\mu = \mu_c$ or $b=b_c$.
We propose to extend the asymptotes, somewhat arbitrarily,
to the very point $\mu = \mu_c$ by replacing
\begin{eqnarray}
    H &  \to & \sqrt{\left(1 - {b \over b_c}\right)^2+ \gamma^2},~~~
   S  \to  \sin \left(\pi - {\pi b \over b_c} \right)+
               \pi \gamma,
\nonumber \\
   u_1 & \to & \sqrt{\left(1 -{\mu \over \mu_c}\right)^2 +{\gamma^2 \over 4}}
\label{HS}
\end{eqnarray}
in Eqs.\ (\ref{high_mu}) and
Eq.\ (\ref{low_mu}). These replacements do not affect significantly
$Q_{\rm syn}$ outside the vicinity of
$\mu = \mu_c$ but produce physically reasonable interpolation
within this vicinity. By matching the modified asymptotes at
$\mu = \mu_c$ and implying again $b \gg 1$ we get
%
   $ \gamma = 1.905 t^{1/2} b^{-1/4} $.
%
In this way we obtain a complete set of equations to describe
$Q_{\rm syn}$ at $\mu < \sqrt{1+2b}$.


%

\newpage

\end{document}